\documentstyle[12pt]{article}
\begin{document}

\title{Domain Growth and Finite-Size-Scaling in the Kinetic Ising
Model}
\author{\\Nigel B.  Wilding\thanks{Present address, Institut f\"u{r}
Physik, Johannes Gutenberg Universit\"{a}t Mainz, Staudinger Weg 7,
D-55128 Mainz, Germany.}, Christian M\"{u}nkel and Dieter W.
Heermann
\\
{\small Institut f\"{u}r Theoretische Physik, Philosophenweg 19,}\\
{\small Universit\"{a}t Heidelberg, D-69120 Heidelberg, Germany}\\
{\small and}\\
{\small Interdisziplin\"{a}res Zentrum f\"{u}r Wissenschaftliches Rechnen}\\
{\small der Universit\"{a}t Heidelberg, D-69120 Heidelberg, Germany}}

\date{}
\setcounter{page}{0}
\maketitle

\begin{abstract}

This paper describes the application of finite-size scaling concepts
to domain growth in systems with a non-conserved order parameter. A
finite-size scaling ansatz for the time-dependent order parameter
distribution function is proposed, and tested with extensive
Monte-Carlo simulations of domain growth in the 2-D spin-flip kinetic
Ising model.  The scaling properties of the distribution functions
serve to elucidate the configurational self-similarity that underlies
the dynamic scaling picture.  Moreover, it is demonstrated that the
application of finite-size-scaling techniques facilitates the accurate
determination of the bulk growth exponent even in the presence of
strong finite-size effects, the scale and character of which are
graphically exposed by the order parameter distribution function.  In
addition it is found that one commonly used measure of domain size--the
scaled second moment of the magnetisation distribution--belies the
full extent of these finite-size effects.

\end{abstract}

\thispagestyle{empty}

\section{Introduction}

When a system in an initially disordered high temperature state is rapidly
quenched to below its transition point, it orders kinetically \cite{REVIEW}.
Small domains of the new equilibrium phase form and grow self-similarly via
the motion of their interfaces.  It is now well established that such growth
processes involve a characteristic time-dependent length ${\overline R}(t)$
-- the `domain-scale'-- identifiable physically as the average domain size.
Moreover, experiments and computer simulations indicate that ${\overline
R}(t)$ satisfies power law behaviour of the form ${\overline R}(t) \sim
t^x$, where $x$ is known as the growth exponent.  Interestingly however, it
transpires that the value of $x$ exhibits a remarkably degree of
insensitivity to the microscopic description of the system, depending only
on whether or not the order parameter is a conserved quantity.  For systems
having a non-conserved order parameter (NCOP)\cite{LIFSHITZ,ALLEN} one finds
$x=\frac{1}{2}$, while in the conserved case (COP)\cite{LIFSHITZ1,BRAY} of
phase separation and spinodal decomposition, it now seems quite generally
agreed that $x=\frac{1}{3}$.

The existence of whole classes of ostensibly quite disparate physical
systems that nevertheless exhibit the {\em same} power law growth is, in
many respects, reminiscent of the universality classes of critical
phenomena.  Indeed the similarities between the two fields run rather deep.
If one regards the divergence of the domain scale ${\overline R}(t)$ at late
times as
analogous to the divergence of the thermal correlation length, then many of
the theoretical tools and concepts familiar from critical phenomena also
carry over (albeit with some important modifications and differences) to the
domain growth context.  The concepts of scaling, fixed points and
renormalisation group flow processes \cite{BRAY,LAI,ZHANG,MAZENKO,MAZENKO2}
underpin the contemporary theoretical view of domain growth kinetics, just
as they do in critical phenomena.

Notwithstanding these advances however, it is probably fair to say that the
present theoretical understanding of domain growth is neither as advanced
nor on as firm a footing as critical phenomena.  This in part stems from a
dearth of exactly soluble models for domain growth, and the consequent lack
of an effective test-bed for new theories and conjectures.  In view of this
situation, it has been common practice to appeal to computer simulation
techniques in order to gain physical insight into domain growth processes.
Indeed, many such simulations studies have been undertaken, of which perhaps
the majority have concentrated on determining the growth law for the system
of interest.  However other studies have drawn their inspiration from more
sophisticated techniques already tried and tested in the context of critical
phenomena.  One such example is the Monte-Carlo renormalisation group
method, which has been successfully adapted to the study of dynamic scaling
in both the COP and NCOP cases \cite{ZHANG,ROLAND,MAZENKO1,KUMAR}.

Given the prevalence of computer simulation studies in the field of
domain growth and the clear parallels with critical phenomena, it
somewhat surprising that hitherto relatively little effort has been
directed towards the application of finite-size-scaling (FSS) ideas to
dynamic scaling phenomena. Indeed little attention seems to have been
given at all to the problems of finite-size effects in simulation
studies of domain growth. By contrast, FSS concepts have for years enjoyed
considerable success in simulation studies of critical phenomena
helping both to mitigate finite size effects, and illuminate the
nature of configurational self-similarity.  Nevertheless, we are aware
of only three papers that consider their utility in the domain growth
context.  We discuss them in turn.

FSS concepts were first applied to domain growth simulations by Vi\~{n}als
and Jasnow \cite{VINALS2}.  These authors studied the time-dependent
structure factor of the kinetic Ising model following quenches to
temperatures well below $T_c$.  A FSS expression for the structure factor
was proposed, and tested using Monte-Carlo simulations for both COP and
NCOP.  Good scaling was observed in the NCOP case and a growth exponent
consistent with $x=\frac{1}{2}$ was obtained.  However the results for the
COP case were somewhat less conclusive, possibly reflecting poor statistics
and the rather small size of the lattice used.

In a related analytical study of the time-dependent Ginsburg-Landau model,
Guo {\em et al} \cite{GUO} derived an approximate expression for the FSS
properties of the structure factor in the NCOP case.  This expression
yielded results in qualitative agreement with the earlier simulation studies
of Vi\~{n}als and Jasnow.  The authors also considered the cross-over from
bulk behaviour to the region where finite-size effects are significant,
pointing out that a fuller understanding of the latter regime requires a
clearer elucidation of the role of metastable strip configurations.

In a theoretical analysis of domain growth following a quench into the
critical region, Milchev {\em et al} \cite{MILCHEV} considered the time
dependent scaling properties of the order parameter distribution function
for NCOP.  By generalising well known expressions for the order parameter
distribution in the context of static critical phenomena, time-dependent FSS
expressions were developed for quenches to $T_c$, as well as to temperatures
slightly below and above $T_c$.  This work also led to a prediction that
critical-slowing-down engenders a weak critical singularity in the
proportionality constant (the so-called `rate') of the growth law
${\overline R}(t) = Bt^x$, of the form $B\propto (1-T/T_c)^{\nu(z-1/x)}$.
However no explicit simulation test of either this prediction or the general
FSS proposals was carried out.

In the present paper we broaden the application of FSS ideas to domain
growth simulations by studying the time-dependent order parameter
distribution function of the 2D spin-flip kinetic Ising (SFKI) model
for NCOP.  The essential features of our study are as follows.  We
have performed extensive Monte-Carlo simulations of domain growth in
the 2D SFKI model following a quench from infinite temperature to
points well within the two phase region.  We obtain the time-dependent
magnetisation distribution function, the structure of which we analyse
within a FSS framework.  Our analysis bears out the configurational
self similarity that lies at the core of the dynamic scaling
hypothesis, Moreover, we show that in contrast to the MCRG method,
application of FSS techniques facilitates accurate determination of
the bulk scaling exponent, even in the presence of strong finite-size
effects. The scale and character of these finite-size effects are
graphically exposed by our distribution functions.

In addition to the FSS analysis, we have also studied in detail one
commonly used measure of domain size, the scaled second moment of the
magnetisation distribution $R^2_M(t) \equiv L^d<M^2(t)>$.  We show
that without a concomitant analysis of configurational structure, this
quantity conveys a misleading impression of the extent of finite size
effects.  We also probe the approach to $t^\frac{1}{2}$ scaling in
$R^2_M(t)$ .  It is demonstrated that a substantial transitory period
precedes the onset of true $t^\frac{1}{2}$ scaling.  The duration of
this `time-lag' is shown to be rather strongly temperature dependent.
A strong temperature dependence is also observed in the
proportionality constant of the growth law, in agreement with earlier
studies.

\section{Background}
\label{sec:bg}

Lifshitz\cite{LIFSHITZ}, and later Cahn and Allen \cite{ALLEN} proposed a
growth equation for NCOP systems on the basis of a phenomenological theory
for the Langevin equation associated with a time-dependent Ginzburg-Landau
model in the absence of noise.  Their principal result, the
Lifshitz-Cahn-Allen (LCA) growth law, is expressed as:

\begin{equation}
{\overline R}(t) =Bt^\frac{1}{2}
\label{eq:lca}
\end{equation}
and is valid for late times. The system-specific proportionality constant $B$
is (within the framework of the LCA theory) temperature independent.

The LCA theory has been tested in numerous experimental \cite{EXPTS} and
computer simulation studies
\cite{MILCHEV,GAWLINSKI,HUMAYUN,SADIQ,FICHTHORN,KASKI}.  Many of these
simulation studies have concentrated on obtaining the growth law for the
system of interest.  This is usually achieved by direct measurements of the
domain size as a function of time.  For this purpose a variety of different
measures of domain size can be employed, including the first and second
moments of the structure factor, the inverse perimeter length and the scaled
second moment of the magnetisation distribution.  A discussion of these
quantities and some of their relative merits can be found in Gawlinski {\em
et al} \cite{GAWLINSKI}.  Having chosen a measure of domain size, the
strategy is then to perform averages over a number of independent quenches,
typically numbering up to a few hundred.  By plotting the results against
$t^x$, power law scaling can be verified and the growth exponent can be
evaluated.  Almost without exception, such measurements of domain growth
yield values for the growth exponent consistent with the LCA prediction
$x=\frac{1}{2}$.

Nevertheless, not all aspects of the LCA theory are in accord with
observation.  Specifically, the neglect of noise in the theory leads to a
temperature independent growth law that is at variance with the
observations of a number of simulation studies
\cite{FICHTHORN,KASKI,SAFRAN,SAHNI}.  These demonstrate that temperature
{\em does} have a significant influence on domain growth, even well away
from the critical point.  In particular it is found that although
$x=\frac{1}{2}$ appears to hold for all sub-critical temperatures, the
magnitude of the proportionality constant $B$ in the growth law decreases
with increasing temperature, the effect being particularly pronounced for
temperatures above about $0.6T_c$.  Current theories suggest that this
`slowing-down' of domain growth is largely attributable to the
disruption of the smooth interface curvature by thermal roughening.  Indeed
a reexamination of the LCA theory, this time including noise \cite{GRANT},
appears to produce results in qualitative agreement with simulation results.
In view of these findings equation~\ref{eq:lca} should be modified to read

\begin{equation}
{\overline R}(t,T)=B(T)t^\frac{1}{2}
\label{eq:newlca}
\end{equation}

Although direct measurements of the growth exponent from the time dependence
of the domain size provide reassuring evidence in support of the dynamic
scaling ideas, they represent only a portion of the information potentially
available from a simulation.  A more stringent and indeed more
physically-compelling test of the dynamic scaling hypothesis, is to directly
observe the time-dependent statistical self-similarity of the coarse-grained
{\em configurational structure}.  In the NCOP case, the essential character
of this structure is uniquely prescribed by the distribution function of the
order parameter, the time dependence of which implicitly embodies the growth
law.  Studies of this quantity therefore represent a potentially exacting
test of the dynamic scaling hypothesis, one that transcends measurements of
the growth exponent alone.  In what follows we develop a FSS approach to the
study of the time dependent order parameter distribution.

\section{Method}
\label{sec:meth}

Let us denote the time dependent distribution of the magnetisation $M$
as $P_L(M,t)$, where $L$ is the linear extent of the system and $t$ is
the time.  Motivated by the work of Milchev {\em et al}
\cite{MILCHEV}, we make the following finite-size-scaling assumption
valid for quenches to temperatures $T < T_c$:

\begin{equation}
P_L(M,t) \simeq  a(L)\tilde P_1(a(L)M,{\overline R}(t) /L,\xi/L)
\label{eq:fss1}
\end{equation}
Here ${\tilde P_1}$ is a scaling function that is expected to be universal
modulo the choice of boundary conditions, $a(L)$ is a system specific
scale factor, ${\overline R}(t)$ is the domain scale and $\xi$ is the thermal
correlation
length.

Now, if the quench temperature lies well below the critical point, $\xi$
will be negligibly small on the scale of $L$, and no critical scaling
behaviour will be apparent.  The only remaining influence of a finite
temperature will be its effect on the growth law prefactor $B(T)$, which is
significant even well away from the critical region \cite{GRANT}.  It is
therefore convenient to simply drop the third argument in
equation~\ref{eq:fss1} and absorb all the non-critical temperature
dependence into ${\overline R}(t)$ to write:

\begin{equation}
P_L(M,t) \simeq\tilde P(M,{\overline R}(t,T) /L)
\label{eq:fss2}
\end{equation}
Equation~\ref{eq:fss2} forms the platform upon which our analysis
procedure rests.  It expresses the basic assertion of the dynamic
scaling hypothesis: the spectrum of coarse-grained configurations is
expected to be invariant under appropriate rescalings of both length
and time.  We note that if one accepts {\em a-priori} the validity of
the dynamic scaling hypothesis, then equation~\ref{eq:fss2}
constitutes a {\em definition} of the domain scale ${\overline
R}(t,T)$ .  We shall exploit this latter feature to make a detailed
evaluation of the growth law within the FSS context. We point out also that
insofar as we obtain the scaling behaviour of the whole distribution
$P_L(M,t)$, encompassing configurational structure on {\em all} length
scales, our study goes well beyond that of Vi\~{n}als and Jasnow
\protect\cite{VINALS2} who only studied the scaling of the structure
factor at three distinct values of wavenumber and did not consider
explicitly the nature of finite-size effects or their role within the
FSS context.

The essential character of the distribution function $P_L(M)$ is
conveyed through the dimensionless fourth order cumulant ratio $G_L$.
This quantity is defined in terms of the moments of the magnetisation
distribution:

\begin{equation}
G_L \equiv \frac{3}{2}-\frac{<M^4>}{2<M^2>^2}
\label{eq:cumrat}
\end{equation}
the value of which ranges from zero for a gaussian distribution to unity for
a pair of delta functions.  The cumulant ratio will prove useful for
analysing the scaling properties of the distribution function.

In the course of the simulation work, we shall also consider one commonly
used measure of domain size, the scaled second moment of the density
distribution \cite{SADIQ}:

\begin{equation}
R^2_M(t) \equiv L^d<M^2(t)>
\label{eq:domsize}
\end{equation}
with $d=2$ in the simulations to be described below.  This quantity is
formally equivalent to the zero wavevector value of the structure factor.
Its behaviour will be compared with the results from the FSS analysis of the
magnetisation distribution.

\section{Monte-Carlo Studies}

\subsection{Computational details}

The Monte-Carlo simulations \cite{BINDER} reported here were all performed
using a Glauber algorithm.  Four system sizes were studied having linear
extent $L=32, 64, 128$ and $256$.  Following convention, periodic boundary
conditions were employed throughout.  In order to avoid spurious `marching'
effects associated with sequential `typewriter' updating, update sites were
visited randomly \cite{GAWLINSKI}.  In this paper, time is expressed in
units of Monte-Carlo steps (MCS) and the update of a single spin is taken as
$1/L^2$ of one MCS.

The simulations were implemented on the Parsytec `Supercluster' Transputer
array of the Interdisziplin\"{a}res Zentrum f\"{u}r Wissenschaftliches
Rechnen der Universit\"{a}t Heidelberg.  Copies of the program were run on
each of $64$ Transputers.  This simple but effective form of parallelism
considerably enhances the rate of data acquisition, allowing the
accumulation of very high statistics.

The time-dependent magnetisation distribution $P_L(M,t)$ was obtained
initially in the form of a histogram recording the magnetisation $M$ as a
function of time.  This histogram was built up over many independent
quenches or `runs'.  As a consistency check, both positive and negative
values of the magnetisation were accumulated separately.  Owing to
limitations on Transputer on-chip memory, it was not possible to histogram
separately all the $L^2+1$ magnetisation states.  A binning procedure was
therefore implemented, with the magnetisation states being distributed over
$513$ bins.  In order to preserve the symmetry of the binning process with
respect to zero magnetisation, the $M=0$ state was allocated its own bin.
The other $L^2$ magnetisation states were distributed uniformly over the
remaining $512$ bins.  The weight in the zero magnetisation bin was
subsequently rescaled to take account of the smaller number of contributing
magnetisation states.

The bulk of the simulation work reported here was performed at a reduced
coupling of $J/K_BT=1$, corresponding to a quench temperature $T\approx
0.44T_c$.  For the $L=32, 64, 128$ and $256$ systems, respective observation
periods of $500, 1000, 1300$ and $2000$ MCS were utilised.  The number of
runs performed in each case was $4\times 10^5, 1.3\times 10^5, 6.4 \times
10^4$ and $5.5\times 10^4$ respectively.  We note that these statistics far
exceed those of previously reported studies by at least an order of
magnitude.

\subsection{Finite-size-scaling results}
\label{sec:fss}

In figure~\ref{fig:distpics} we present the time evolution of the
magnetisation distribution function $P_L(M,t)$ for the $L=32$ and $L=128$
system sizes.  The functional form of these distributions serves to
elucidate the nature of the ordering processes in the system.  The salient
features are best illustrated by considering the form of the distribution
function at a number of discrete timeslices.  For convenience we shall
employ the $L=32$ data for this purpose, but our remarks apply equally to
all system sizes.

Figure~\ref{fig:slices} shows the $L=32$ magnetisation distribution at times
$t = 20, 50, 75, 100, 175, 256,$ $512$ MCS.  Clearly at very early times,
the distribution is sharply peaked around $M=0$, reflecting the fact that
the system is still highly disordered and only small domains have begun to
form.  As time progresses however, and the domains grow ever larger on the
scale of the system, the distribution spreads out and weight is transferred
from the central portion towards the periphery.  Interestingly though, this
broadening of the distribution is not uniform as one might expect.  Instead
the distribution starts to develop quite early on into a {\em three peaked}
structure.  We discuss the origin of each of these peaks in turn.

The first peak is centred on $M=0$ and represents metastable strip
configurations.  These strip configurations comprise a single domain which
spans the system along one of the lattice directions.  At the temperature
studied, strip configurations start to appear for $t^\frac{1}{2}\simeq
0.2L$.  Since the strip interface is essentially flat, there is no driving
force for further growth, and decay can only proceed by thermal fluctuations
\cite{SAFRAN}.  Consequently strip domains are extremely long-lived, being
essentially stable on the timescale of the other growth processes we
consider.

The second peak, which we term the `dynamic peak', corresponds to
configurations that are still growing, i.e.  those that have not evolved
into strip configurations.  An example of a dynamic configuration is a
circular domain embedded in a `sea' of the opposite phase.  At short times,
the dynamic peak is to be found near $M=0$ where it is superimposed on the
central peak of metastable configurations.  As time progresses however and
the domains grow, the dynamic peak `moves' to successively larger $M$ values
where it can be resolved from the metastable peak.  Since metastable states
are manufactured from dynamic ones, strip configurations necessarily have
magnetisation values that lie under the dynamic peak at their moment of
formation.  We find that in approximately $35\%$ of runs, the system evolves
into a metastable strip state.

The third peak, which we term the `saturation peak', represents equilibrium
configurations which start to appear at approximately the same time as strip
configurations.  This peak is very narrow (its width varies like $L^{-d}$)
and for low temperatures is situated very close to $M=\pm1$.  The saturation
peak draws its weight from the large-$M$ tail of the dynamic peak.  All
those surviving dynamic configurations that escape entrapment as metastable
strips, eventually become saturated.  Thus for late times (though still
short compared to the characteristic lifetime of metastable strips), the
system enters a quasi-asymptotic regime where all growth has effectively
ceased and the distribution function comprises only the central metastable
peak and the saturation peak.  It is to be expected that the
quasi-asymptotic form of the distribution function is $L$-independent.

We turn now to an examination of the time-dependent configurational self
similarity that underlies the dynamic scaling picture.  Specifically we
consider the transformations (rescalings of time) required to perform the
data collapse of distribution functions for different $L$.  In
figure~\ref{fig:datcol}, we show example timeslices of the distribution
function $P_L(M,t)$ for various choices of $L$ and $t$.  The data correspond
to three distinct values of $z\equiv t^\frac{1}{2}/L$ (cf.
equations~\ref{eq:newlca} and \ref{eq:fss2}), namely $z\approx 0.17, 0.28$
and $0.31$.  In each instance the timeslices shown were chosen specifically
to optimise the quality of the data collapse as quantified by the matching
of the cumulant ratio.  Figure~\ref{fig:datcol}(a) shows the case $z \approx
0.17$ for the four system sizes $L=32, 64, 128$ and $256$, at respective
times $t=25, 110, 480, 1952$.  The case $z\approx 0.28$ is shown in
figure~\ref{fig:datcol}(b) for the $L=64$ and $L=128$ systems at respective
times $t=313$ and $1300$.  Finally, figure~\ref{fig:datcol}(c) shows the
data for $z \approx 0.31$ for system sizes $L=32$ and $L=64$, at times
$t=100$ and $t=420$ respectively.

In each of the examples shown in figure~\ref{fig:datcol}(a)--(c), the
quality of the data collapse is high, although the extent of finite-size
effects differs markedly between the three cases.  In case (a) there is
nothing in the structure of the distribution functions to suggest that
finite-size effects are yet significant.  On the other hand, for cases (b)
and (c) there is clearly a high proportion of static (metastable and
saturated) configurations, as evidenced by the triple-peaked structure of
the distribution function.  We term this latter regime the strongly
finite-size scaling region.  At first sight, the observation of
time-dependent scaling even in the presence of static configurations, is
somewhat counter-intuitive.  We wish to argue however, that this is actually
a {\em feature} of our finite-size-scaling approach.

To this end it is instructive to consider the nature of typical
configurations in the strongly finite-size region $(z > 0.2)$.  In this
regime one finds both dynamic configurations and static configurations, the
former of which grow like $t^\frac{1}{2}$ while the latter, of course, don't
grow at all.  The crucial point underpinning the observation of scaling
behaviour, is that metastable or saturated configurations are {\em only}
formed when a dynamic configuration spans the system.  Thus their formation
rate at a given time $t$ (and hence their magnetisation distribution) is
entirely mediated by the distribution of dynamic clusters which barely span
the system at time $t$.  Now, since the distribution of these
barely-spanning dynamic clusters is controlled by the scaling variable
${\overline R}(t,T)/L$, it follows that the {\em whole distribution}
(comprising both dynamic {\em and} static contributions) also scales with
this variable.  Accordingly $t^\frac{1}{2}$ FSS behaviour is observed even
at times when a significant proportion of configurations have stopped
growing.  Moreover, measurements of the growth exponent $x$ extracted from
the scaling behaviour of the distribution functions are expected to be
reliable at times long after the simple picture of bulk growth (with an
associated physically well-defined domain size) has broken down.  This is,
we feel, analogous to the situation in critical phenomena, where application
of FSS techniques permit accurate measurements of critical exponents even
when the correlation length exceeds the system size.

The data collapse shown in figure~\ref{fig:datcol}, gives particular
examples of dynamic scaling.  However in order to demonstrate the full
extent of the scaling behaviour that resides in our distributions, it is
expedient to consider the time-dependence of the cumulant ratio.  Figure
{}~\ref{fig:cumrat}(a) shows the measured cumulant ratio (as prescribed in
equation~\ref{eq:cumrat}) expressed as a function of time for all four of
the system sizes studied.  Before examining the cumulant scaling behaviour
however, we should point out two general features of the data.  Firstly at
short times the data for $L=128$ and $L=256$ appears somewhat noisy.  This
effect is however simply an artifact of our binning procedure and reflects
the fact that for early times (and especially for larger system sizes), the
distribution is very narrow and contributes significant weight to only a
small number of bins.  Measurements of the cumulant ratio from data
distributed over so few bins exhibit an enhanced degree of sensitivity to
small statistical deviations from symmetry.  Secondly, at late times the
data for the $L=32$ system shows the cumulant ratio appearing to saturate at
a value $G_L\approx 0 .  8$, well short of its true asymptotic value of
unity.  This leveling-off of $G_L$ reflects the cessation of growth that
occurs in the quasi-asymptotic regime.  The discrepancy in the cumulant
value signals the presence of metastable strip configurations.

In order to expose the time-dependent scaling behaviour of the cumulant
ratio, we have reexpressed the data of figure~\ref{fig:cumrat}(a) in terms
of the variable $z\equiv t^\frac{1}{2}/L$.  The rescaled results are shown
in figure~\ref{fig:cumrat}(b).  Except at short times, where as already
noted, measurements of the cumulant are effected by noise, the curves are to
a good approximation parallel to one another (there is some deviation for
the $L=32$ system, and these are addressed in the discussion section).  We
note that this parallelism extends well beyond $z=0.2$, thus corroborating
our assertion that $P_L(M,t)$ manifests $t^\frac{1}{2}$ scaling even in the
presence of strong finite-size effects.  Interestingly, however although the
curves for $L=128$ and $L=256$ coincide for late times, the data for $L=32$
and $L=64$ are displaced from the limiting (large $L$) case by an amount
that decreases with increasing $L$.  To effect coincidence and achieve
scaling for the smaller system sizes, it is therefore necessary to make an
empirical redefinition of the domain scale ${\overline R}(t,T)$ to include
an additive $L$-dependent term viz:

\begin{equation}
{\overline R}(t,T,L) = A(L)+B(T)t^\frac{1}{2}
\end{equation}
where $A(L)$ is a system specific quantity.  Use of this modified form of
the domain scale in the scaling variable $z\equiv {\overline R}/L$ allows
the scaled cumulant data to be brought into coincidence for all system
sizes.  Specifically, we find that the $L=32$ and $L=64$ data can be brought
into correspondence with the $L=128$ and $L=256$ data by choices
$A(32)\approx 0.7, A(64)\approx 0.8$.

\subsection{The scaled second moment:  $R^2_M(t)\equiv L^d<M^2(t)>$}

One commonly employed measure of the domain size (actually the {\em square}
of the domain size) in simulation studies of NCOP growth, is the scaled
second moment of the magnetisation, $R^2_M(t) \equiv L^d<M^2(t)>$
\cite{SADIQ}.  Although a number of simulation studies of this quantity have
been carried out, it is in practice rather difficult to obtain highly
accurate measurements due to its lack of self averaging \cite{MILCHEV}.
This implies that the relative statistical error cannot be reduced by
increasing the system size, but only by performing a very large number of
runs--a task that is computationally intensive.  Perhaps as a consequence,
none of the previous studies of $R^2_M(t)$ have provided very detailed
information on its behaviour.

In the present study, $R^2_M(t)$ is obtained as a {\em by-product} of our
measurements of $P_L(M,t)$ .  Moreover owing to the large number of runs
needed to accumulate our magnetisation distribution functions, we have been
able to obtain very precise estimates of its behaviour.  These estimates
were extracted directly from the measured second moment of the magnetisation
distributions and are shown in figure~\ref{fig:domsize} expressed as a
function of time.  In contrast to the scaled cumulant ratios, the $R^2_M(t)$
data for all systems sizes coincide to a very high degree of accuracy.
Furthermore the data also appears linear in time, as indeed one would expect
on the basis of the LCA growth law.  The data remains linear until
$t^\frac{1}{2} \approx 0.4L$ at which point the curves tail off.

It transpires on closer inspection, however, that the data is not as linear
as a cursory examination would suggest.  In figure~\ref{fig:approach} we
show the data of figure~\ref{fig:domsize} with the $t$ dependence divided
out.  This clearly shows that $t^\frac{1}{2}$ scaling does not in fact set
in until after an initial transitory period---approximately $300$ MCS are
required for $R^2_M(t) /t$ to attain $95\%$ of its late-time value.  This
`time-lag' manifests the fact that scaling behaviour only becomes apparent
once the domains have grown large on the scale of the largest microscopic
length (in this case the lattice spacing).  We note additionally that for
the $L=32$ there is no real `window' of $t^\frac{1}{2}$ scaling at all since
all growth ceases well before $300$ MCS.

An investigation of the influence of temperature on domain growth has also
been performed.  Measurements of $R^2_M(t)$ at short times ($t < 300$ MCS)
were made for the $L=128$ system at a variety of temperatures between $T=0$
and $T=T_c$.  For each temperature studied, $2.5\times 10^4$ runs were
performed.  The results, presented in figure~\ref{fig:tempeff}(a), clearly
show that domain growth slows as the temperature is increased, in accord
with earlier findings \cite{FICHTHORN,KASKI,SAFRAN,SAHNI}.  In order to
quantify the degree of this slowing-down, we have obtained estimates for the
square of the proportionality constant $B(T)$ of the growth law.  These
estimates, representing the gradient of the $R^2_M(t)$ curve at $t=300$ MCS,
are plotted as a function of temperature in figure~\ref{fig:tempeff}(b).
Also shown is the theoretical expression of Grant and Gunton \cite{GRANT},
the arbitrary vertical scale of which has been chosen so as to match the
simulation data at $T=0$.  The observed behaviour is in qualitative
agreement with the theoretical prediction, although the theory substantially
underestimates the temperature at which the growth rate vanishes.

Finally, the effect of temperature on the approach to scaling has also
been examined, by dividing out the time dependence.  The results,
figure~\ref{fig:tempeff}(c), demonstrate that the approach to scaling
is more protracted at higher temperatures, presumably reflecting the
slower growth rate.  Nevertheless, all indications are that
$t^\frac{1}{2}$ scaling is still obeyed for late times. We note
however, that the strength of the roughening effect in the
near-critical region will almost certainly mask the proposed weak
critical singularity of the growth law proportionality constant
\protect\cite{MILCHEV}, complicating any verification of its existence
by computer simulation methods.

\section{Discussion}

In this paper we have explored the application of finite-size-scaling
techniques to the domain growth problem.  Our results for the matching
of the distribution function $P_L(M,t)$, as expressed through our
scaled cumulant data, bear out the time-dependent configurational
self-similarity that underlies the dynamic scaling
hypothesis. Moreover, we find that when analysed within a FSS
framework, the distributions yield accurate estimates for the bulk
scaling exponent $n=1/2$, even in the regime where the system is
strongly afflicted by finite size effects. Clearly therefore the
present method should prove useful in circumventing the limitations
imposed by finite computer resources in studies of domain growth.

We have also examined in detail the behaviour of the scaled second
moment of the magnetisation $R^2_M(t)\equiv L^d<M^2(t)>$, used as a
measure of the domain size.  Although this quantity exhibits
$t^\frac{1}{2}$ scaling for late times, our results shows that
appreciable deviations from this behaviour persist until at least
$300$ MCS.  Indeed for the $L=32$ case studied here, the system never
attains the asymptotic scaling regime because all growth stops well
before $300$ MCS.  This effect is presumably responsible for the
observation (noted in section~\ref{sec:fss}), that the scaled cumulant
value for the $L=32$ system is not quite parallel to those of the
larger system sizes (cf.  figure~\ref{fig:cumrat}(b)).  Evidently
therefore, system sizes of at least $L=64$ must be employed if
anything approaching `true' $t^\frac{1}{2}$ scaling is to be observed
at all.

In this context it is also interesting to note that $R^2_M(t)$
exhibits few of the finite-size effects manifested by the
distributions themselves.  Thus for instance, the data of
figure~\ref{fig:domsize} coincide for all system sizes, in contrast to
the behaviour of the scaled cumulant ratio of
figure~\ref{fig:cumrat}(b).  Also, $R^2_M(t)$ is apparently
insensitive to the presence of metastable strip and saturated
configurations, remaining stubbornly linear (at the temperature
studied) up to $t^\frac{1}{2}\approx 0.4L$, corresponding to
$R_M=0.65L$.  In contrast, the evidence of our distributions shows
that metastable and saturated configurations actually appear at
substantially earlier times than this, being already evident for
$t^\frac{1}{2} = 0.2L$, corresponding to $R_M=0.32L$.

Clearly therefore caution should be exercised when attempting to gauge
the extent of finite-size effects without an accompanying analysis of
configurational structure.  Particular care is called for when
performing Monte Carlo renormalisation group (MCRG) studies of domain
growth.  In such studies one typically applies a local blocking
transformation, the action of which is intended to produce
configurations whose statistical properties are appropriate to an
earlier time.  Evidently however such a procedure must fail when
applied to metastable or saturated configurations because repeated
applications of a blocking transformation cannot produce a
non-spanning domain.  The MCRG method is therefore strictly only
applicable in the effectively bulk regime where no metastable or
saturated configuration have yet evolved. In this respect, the present
FSS method possesses a major advantage over the MCRG approach.
The size of our coarse-graining length is simply that of the system
itself, and the effect of changing this length is studied not by a
blocking transformation, but by comparison with {\em independent}
simulations of different sizes.  Consequently the difficulties
associated with blocking transformations are circumvented.

Finally with regard to extensions of the present work, it has been
suggested that the appearance of metastable strip configurations may
merely be an artifact of the use of periodic boundary conditions
\cite{GAWLINSKI,VINALS1}.  It would now be of interest to explore in
detail the degree to which the boundary conditions effect the nature
and extent of finite-size effects, particularly with regard to the
stability of strip configurations.  We intend to report on such
extensions in future work.


\subsection*{Acknowledgments}

NBW and Ch.M acknowledge financial support from the Graduiertenkolleg f\"{u}r
Modellierung und wissenschaftliches Rechnen in Mathematik und
Naturwissenschaften, IWR Universit\"{a}t Heidelberg.

\newpage
\pagestyle{empty}

\begin{figure}[h]

\caption{The time evolution of (a) the $L=32$, and (b) the $L=128$
magnetisation distribution functions $P_L(M,t)$ for times in the range
$0-512$ MCS and $0-1300$ MCS respectively.  The distribution at each instant
is normalised to unit integrated weight.  In (a) the saturation peaks at
$M\simeq\pm 1$ have been truncated in order to accentuate the central
structure.}

\label{fig:distpics}
\end{figure}

\begin{figure}[h]

\caption{Timeslices of the $L=32$ magnetisation distribution $P_L(M,t)$ at
times $t=20, 50, 75, 100, 175,$ $256, 512$ MCS.  All distributions are
normalised to unit integrated weight.}

\label{fig:slices}
\end{figure}

\begin{figure}[h]

\caption{The data collapse of the normalised magnetisation distribution
$P_L(M)$ for various system sizes and times.  (a) $L=32,64, 128$ and $256$
at times $t=25, 110, 480$ and $1952$ MCS respectively.  (b) $L=64$ and
$L=128$ at times $t=313$ and $t=1300$ MCS respectively.  (c) $L=32$ and
$L=64$ at times $t=100$ and $t=420$ MCS respectively}

\label{fig:datcol}
\end{figure}

\begin{figure}[h]

\caption{(a) The cumulant ratio $G_L$ of the magnetisation distribution,
prescribed in equation~\protect\ref{eq:cumrat}, and expressed as a function
of time.  The data shown corresponds to the four system sizes $L=32, 64,
128$ and $256$.  (b) The data of (a) reexpressed as a function of the
variable $z=t^\frac{1}{2}/L$.  In the interests of clarity, only data for $0
< z < 0.4$ is shown and the data points have been suppressed.  The full data
set is shown in the inset.}

\label{fig:cumrat}
\end{figure}

\begin{figure}[h]


\caption{(a) The quantity $R^2_M(t)$, defined in the text, and expressed as
a function of time.  Data is shown for four system sizes $L=32, 64, 128$ and
$256$.  Errors are less than 0.5\%.  The data points have been suppressed
for clarity.  (b) Log-log plot of the data of (a).}

\label{fig:domsize}
\end{figure}

\begin{figure}[h]

\caption{The quantity $R^2_M(t)/t$ expressed as a function of time for the
four system sizes $L=32, 64, 128$ and $256$.  }

\label{fig:approach}
\end{figure}

\vspace{-2.3cm}
\begin{figure}[h]

\caption{(a) $R^2_M(t)$ for the $L=128$ system at a selection of
temperatures in the range $0$--$T_c$.  Errors are less than 1\% .  (b) The
gradient of $R^2_M(t)$ at $t=300$ MCS for various values of $T/T_c$.  The
dashed line through the data points serves merely to guide the eye.  Errors
are comparable with symbol sizes.  Also shown (solid line) is the
theoretical prediction of Grant and Gunton \protect\cite{GRANT}, scaled in
the manner described in the text.  (c) The data of (a) with the time
dependence divided out.  }

\label{fig:tempeff}

\end{figure}


\end{document}